\documentstyle[prd,eqsecnum,aps,12pt]{revtex}
\begin{document}
\draft
\title{\noindent{\large\tt {} \hfill DAMTP 94/34}\\ \vskip 1.0cm
Nonlinear interaction between electromagnetic fields at high temperature}
\author{F. T. Brandt, J. Frenkel}
\address{Instituto de F\'\i sica, Universidade de S\~ao Paulo,\\
S\~ao Paulo, 01498 SP, Brasil}
\author{J. C. Taylor}
\address{Department of Applied Mathematics and Theoretical Physics,
University of Cambridge,\\
Cambridge, CB3 9EW, UK}
\maketitle
\vskip 1.0cm
\begin{abstract}
The electron-positron `box' diagram produces an effective action which
is fourth order in the electromagnetic field.
We examine the behaviour of this effective
action at high-temperature (in analytically continued imaginary-time
thermal perturbation theory). We argue that there is a finite, nonzero limit
as $T\rightarrow \infty$ (where $T$ is the temperature). We calculate
this limit in the nonrelativistic static case, and in the long-wavelength
limit. We also briefly discuss the self-energy in 2-dimensional QED, which is
similar in some respects.
\end{abstract}


\section{Introduction}

\label{intro}

The effective action due to the electron-positron `box' diagram was
calculated in 1950 \cite{KN}. It is fourth order in the electromagnetic field.
The object of this paper is to investigate  the
behaviour of the QED effective action at temperatures
high compared to the electron mass and to the frequencies and wave numbers
of the electromagnetic field (see also \cite{DittrichLoewe}).
We use analytically continued imaginary-time
thermal perturbation theory (which is known \cite{Evans,Taylor} to be
related to real-time theory).
The question which we seek to answer is: what is the behaviour at
high-temperature?
Individual contributions contain $T^2$ and $\log(T)$
terms, where $T$ is the temperature ($T^3$ and $T$ terms cancel by symmetry).
However, gauge invariance imposes strong constraints (just as it does at
zero temperature, forbidding UV divergence for example), and one expects
strong cancellations. In fact we are able to confirm explicitly the
cancellation of the $T^2$  terms. Note that the dependance upon $T$
for high $T$
is not necessarily simply related to the UV divergence or
convergence of the zero-temperature amplitude. In QCD, for example,
the $n$-gluon functions, for all $n$, go as $T^2$ although they are UV
finite for $N>4$.

The main concern of this paper is possible $\log(T)$ terms. We have not
managed to demonstrate their absence by explicit computation. Instead we give
an indirect argument. This (Sec. \ref{SEC4})
is based upon first studying the static case and then using Lorentz invariance.

We therefore expect the effective action to have a finite, non-zero limit
as $T \rightarrow \infty$. Like the zero-temperature action,
this must be gauge-invariant; but it may be more
complicated because it is not necessarily Lorentz-invariant. We have
not been able to compute it in general. In two special
cases, the nonrelativisitic static case (which is just a number)
and the long wavelength limit (see the Appendix),
we have been able to compute the
high temperature limit. We have also checked by direct calculation
the cancellation of $\log(T)$ terms in the special case in which all
the wave-vectors are parallel.

There is an example which has some points of resemblance. That is the
self-energy function in 2-dimensional QED. Here (although the zero-temperature
function is UV finite) it is not immediately obvious whether there
is a $\log(T)$ term.
We find, in fact, that there is a finite limit at high $T$.

Four-photons graphs with two thermal photons and non-thermal fermions have
been considered as subdiagrams contributing in higher order to the photon
self energy \cite{TARRACH}. These lead in the low temperature
limit, to temperature-dependent corrections to the dielectric constant and
speed of light behaving like $(T/m)^4$. As a possible application of our
four-photon result, one could thermalise two photons obtaining, following
\cite{TARRACH}, a higher order correction to the photon self-energy at
high temperature.

In thermal QCD, it is known that a study of
the high-temperature behaviour of thermal graphs is an important tool in
resumming perturbation theory. In QED, power dependence on $T$ is confined to
self-energy graphs. The next question to ask is whether there are graphs with
a $\log(T)$ behaviour at large $T$. This paper addresses that problem.

\section{The thermal perturbation theory}

\label{SEC2}

\begin{figure}[t]
\setlength{\unitlength}{0.008000in}%
\begingroup\makeatletter\ifx\SetFigFont\undefined
% extract first six characters in \fmtname
\def\x#1#2#3#4#5#6#7\relax{\def\x{#1#2#3#4#5#6}}%
\expandafter\x\fmtname xxxxxx\relax \def\y{splain}%
\ifx\x\y   % LaTeX or SliTeX?
\gdef\SetFigFont#1#2#3{%
  \ifnum #1<17\tiny\else \ifnum #1<20\small\else
  \ifnum #1<24\normalsize\else \ifnum #1<29\large\else
  \ifnum #1<34\Large\else \ifnum #1<41\LARGE\else
     \huge\fi\fi\fi\fi\fi\fi
  \csname #3\endcsname}%
\else
\gdef\SetFigFont#1#2#3{\begingroup
  \count@#1\relax \ifnum 25<\count@\count@25\fi
  \def\x{\endgroup\@setsize\SetFigFont{#2pt}}%
  \expandafter\x
    \csname \romannumeral\the\count@ pt\expandafter\endcsname
    \csname @\romannumeral\the\count@ pt\endcsname

  \csname #3\endcsname}%
\fi
\fi\endgroup
\begin{picture}(760,232)(60,290)
%
%Feito por Fernando Tadeu C. Brandt em 30/6/94
%
\thicklines
\multiput( 80,480)(3.2,-3.2){13}{\makebox(0.7,1.1){\SetFigFont{10}{12}{rm}.}}
\multiput(360,480)(3.2,-3.2){13}{\makebox(0.7,1.1){\SetFigFont{10}{12}{rm}.}}
\multiput(640,480)(3.2,-3.2){13}{\makebox(0.7,1.1){\SetFigFont{10}{12}{rm}.}}
\multiput(240,480)(-3.2,-3.2){13}{\makebox(0.7,1.1){\SetFigFont{10}{12}{rm}.}}
\multiput(520,480)(-3.2,-3.2){13}{\makebox(0.7,1.1){\SetFigFont{10}{12}{rm}.}}
\multiput(800,480)(-3.2,-3.2){13}{\makebox(0.7,1.1){\SetFigFont{10}{12}{rm}.}}
\multiput( 80,320)(3.2,3.2){13}{\makebox(0.7,1.1){\SetFigFont{10}{12}{rm}.}}
\multiput(360,320)(3.2,3.2){13}{\makebox(0.7,1.1){\SetFigFont{10}{12}{rm}.}}
\multiput(640,320)(3.2,3.2){13}{\makebox(0.7,1.1){\SetFigFont{10}{12}{rm}.}}
\multiput(240,320)(-3.2,3.2){13}{\makebox(0.7,1.1){\SetFigFont{10}{12}{rm}.}}
\multiput(520,320)(-3.2,3.2){13}{\makebox(0.7,1.1){\SetFigFont{10}{12}{rm}.}}
\multiput(800,320)(-3.2,3.2){13}{\makebox(0.7,1.1){\SetFigFont{10}{12}{rm}.}}
\put(120,360){\vector( 1, 0){ 45}}
\put(165,360){\line( 1, 0){ 35}}
\put(200,360){\vector( 0, 1){ 45}}
\put(200,405){\line( 0, 1){ 35}}
\put(200,440){\vector(-1, 0){ 45}}
\put(155,440){\line(-1, 0){ 35}}
\put(120,440){\vector( 0,-1){ 45}}
\put(400,360){\vector( 1, 0){ 45}}
\put(445,360){\line( 1, 0){ 35}}
\put(480,360){\vector( 0, 1){ 45}}
\put(480,405){\line( 0, 1){ 35}}
\put(480,440){\vector(-1, 0){ 45}}
\put(435,440){\line(-1, 0){ 35}}
\put(400,440){\vector( 0,-1){ 45}}
\put(680,360){\vector( 1, 0){ 45}}
\put(725,360){\line( 1, 0){ 35}}
\put(760,360){\vector( 0, 1){ 45}}
\put(760,405){\line( 0, 1){ 35}}
\put(760,440){\vector(-1, 0){ 45}}
\put(715,440){\line(-1, 0){ 35}}
\put(680,440){\vector( 0,-1){ 45}}
\put(120,360){\line( 0, 1){ 35}}
\put(400,360){\line( 0, 1){ 35}}
\put(680,360){\line( 0, 1){ 35}}
\put( 60,500){\makebox(0,0)[lb]{\smash{\SetFigFont{12}{14.4}{it}k}}}
\put( 80,500){\makebox(0,0)[lb]{\smash{\SetFigFont{12}{14.4}{rm},}}}
\put(220,500){\makebox(0,0)[lb]{\smash{\SetFigFont{12}{14.4}{it}k}}}
\put(240,500){\makebox(0,0)[lb]{\smash{\SetFigFont{12}{14.4}{rm},}}}
\put(340,500){\makebox(0,0)[lb]{\smash{\SetFigFont{12}{14.4}{it}k}}}
\put(360,500){\makebox(0,0)[lb]{\smash{\SetFigFont{12}{14.4}{rm},}}}
\put(500,500){\makebox(0,0)[lb]{\smash{\SetFigFont{12}{14.4}{it}k}}}
\put(520,500){\makebox(0,0)[lb]{\smash{\SetFigFont{12}{14.4}{rm},}}}
\put(620,500){\makebox(0,0)[lb]{\smash{\SetFigFont{12}{14.4}{it}k}}}
\put(640,500){\makebox(0,0)[lb]{\smash{\SetFigFont{12}{14.4}{rm},}}}
\put(780,500){\makebox(0,0)[lb]{\smash{\SetFigFont{12}{14.4}{it}k}}}
\put(800,500){\makebox(0,0)[lb]{\smash{\SetFigFont{12}{14.4}{rm},}}}
\put(100,500){\makebox(0,0)[lb]{\smash{\SetFigFont{12}{14.4}{rm}$\sigma$}}}
\put(260,500){\makebox(0,0)[lb]{\smash{\SetFigFont{12}{14.4}{rm}$\lambda$}}}
\put(380,500){\makebox(0,0)[lb]{\smash{\SetFigFont{12}{14.4}{rm}$\sigma$}}}
\put(540,500){\makebox(0,0)[lb]{\smash{\SetFigFont{12}{14.4}{rm}$\nu$}}}
\put(660,500){\makebox(0,0)[lb]{\smash{\SetFigFont{12}{14.4}{rm}$\lambda$}}}
\put(820,500){\makebox(0,0)[lb]{\smash{\SetFigFont{12}{14.4}{rm}$\sigma$}}}
\put( 60,300){\makebox(0,0)[lb]{\smash{\SetFigFont{12}{14.4}{it}k}}}
\put( 80,300){\makebox(0,0)[lb]{\smash{\SetFigFont{12}{14.4}{rm},}}}
\put(100,300){\makebox(0,0)[lb]{\smash{\SetFigFont{12}{14.4}{rm}$\mu$}}}
\put(340,300){\makebox(0,0)[lb]{\smash{\SetFigFont{12}{14.4}{it}k}}}
\put(360,300){\makebox(0,0)[lb]{\smash{\SetFigFont{12}{14.4}{rm},}}}
\put(380,300){\makebox(0,0)[lb]{\smash{\SetFigFont{12}{14.4}{rm}$\mu$}}}
\put(620,300){\makebox(0,0)[lb]{\smash{\SetFigFont{12}{14.4}{it}k}}}
\put(640,300){\makebox(0,0)[lb]{\smash{\SetFigFont{12}{14.4}{rm},}}}
\put(660,300){\makebox(0,0)[lb]{\smash{\SetFigFont{12}{14.4}{rm}$\mu$}}}
\put(780,300){\makebox(0,0)[lb]{\smash{\SetFigFont{12}{14.4}{it}k}}}
\put(800,300){\makebox(0,0)[lb]{\smash{\SetFigFont{12}{14.4}{rm},}}}
\put(820,300){\makebox(0,0)[lb]{\smash{\SetFigFont{12}{14.4}{rm}$\nu$}}}
\put( 70,490){\makebox(0,0)[lb]{\smash{\SetFigFont{8}{8.0}{rm}4}}}
\put(230,490){\makebox(0,0)[lb]{\smash{\SetFigFont{8}{8.0}{rm}3}}}
\put(350,490){\makebox(0,0)[lb]{\smash{\SetFigFont{8}{8.0}{rm}4}}}
\put(510,490){\makebox(0,0)[lb]{\smash{\SetFigFont{8}{8.0}{rm}2}}}
\put(630,490){\makebox(0,0)[lb]{\smash{\SetFigFont{8}{8.0}{rm}3}}}
\put(790,490){\makebox(0,0)[lb]{\smash{\SetFigFont{8}{8.0}{rm}4}}}
\put( 70,290){\makebox(0,0)[lb]{\smash{\SetFigFont{8}{8.0}{rm}1}}}
\put(350,290){\makebox(0,0)[lb]{\smash{\SetFigFont{8}{8.0}{rm}1}}}
\put(630,290){\makebox(0,0)[lb]{\smash{\SetFigFont{8}{8.0}{rm}1}}}
\put(790,290){\makebox(0,0)[lb]{\smash{\SetFigFont{8}{8.0}{rm}2}}}
\put(220,300){\makebox(0,0)[lb]{\smash{\SetFigFont{12}{14.4}{it}k}}}
\put(240,300){\makebox(0,0)[lb]{\smash{\SetFigFont{12}{14.4}{rm},}}}
\put(255,300){\makebox(0,0)[lb]{\smash{\SetFigFont{12}{14.4}{rm}$\nu$}}}
\put(500,300){\makebox(0,0)[lb]{\smash{\SetFigFont{12}{14.4}{it}k}}}
\put(520,300){\makebox(0,0)[lb]{\smash{\SetFigFont{12}{14.4}{rm},}}}
\put(540,300){\makebox(0,0)[lb]{\smash{\SetFigFont{12}{14.4}{rm}$\lambda$}}}
\put(230,290){\makebox(0,0)[lb]{\smash{\SetFigFont{8}{8.0}{rm}2}}}
\put(510,290){\makebox(0,0)[lb]{\smash{\SetFigFont{8}{8.0}{rm}3}}}
\end{picture}

\nopagebreak
\bigskip
\caption[f1]{
\label{f1}{
Box diagrams which contribute to the nonlinear
electromagnetic interaction. Dotted lines represent photons, and
solid lines stand for electrons.}}
\end{figure}
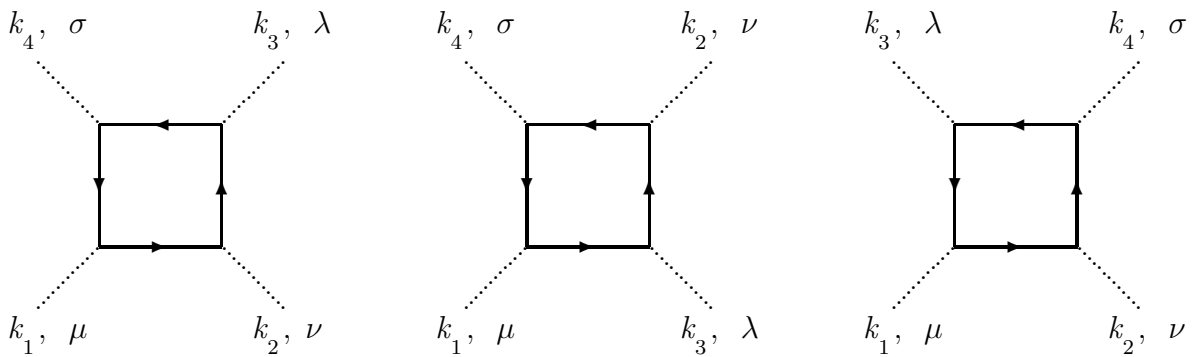

The graphs which contribute to the nonlinear electromagnetic interaction
are shown in Fig. \ref{f1}. There are three other contributions obtained by
charge conjugation.
%%%%%%%%%%%%%%%%%%%%%%%%%%%% Fig  1 %%%%%%%%%%%%%%%%%%%%%%%%%%%%%%%%%%%%%%%%%
%%%%%%%%%%%%%%%%%%%%%%%%%%%%%%%%%%%%%%%%%%%%%%%%%%%%%%%%%%%%%%%%%%%%%%%%%%%%

The analytically continued imaginary-time thermal perturbation
theory can be formulated \cite{Barton,FT} so as to express the thermal
amplitude (having subtracted the zero-temperature part)
in terms of amplitudes for forward scattering of an electron
in an external electromagnetic field, as depicted in Fig. \ref{f2}.
There are 24 Feynman diagrams like this one, which are
obtained by even and odd permutations of the external momenta and
polarizations. The corresponding analytic expression has the form
%%%%%%%%%%%%%%%%%%%%%%%%%%%%%%%%%%%%% EQ. 1
\begin{equation}\label{eq1}
\frac{{ e^4}}{(2\pi )^3} \int_0^{\infty}
\frac{{ q}^2}{{ 2Q^0}} {\rm d} q N({Q^0}) \int d\Omega
\sum_{ijkl}
{ B}_{(ijkl)}^{\mu \nu \lambda \sigma}({ k_1 ,k_2 ,k_3};Q).
\end{equation}
%%%%%%%%%%%%%%%%%%%%%%%%%%%%%%%%%%%%%
Here
${ q}=|{\bf Q}|$,
${Q^0}=({ q^2 +m^2})^{1/2}$, ${ N(Q^0)}=({ 1+{\rm e}^{Q^0/T}})^{-1}$,
$\int d\Omega$ is an integral over the directions of ${\bf Q}$, and the
sum is over the permutation $(ijkl)$ of $(1234)$.
Each ${ B}$ has a numerator which is a
Dirac trace containing a projection operator
$\left({{\bf \slash}\hskip-.65em\relax Q+m}\right)$. For example
%%%%%%%%%%%%%%%%%%%%%%%%%%%%%%%%%%%%%%% EQ. 2
\begin{equation}
\label{B1234}
{ B}_{(1234)}^{\mu\nu\lambda\sigma}=\displaystyle{
\frac{{\rm tr}\left[
\left({{\bf \slash}\hskip-.65em\relax Q+m}\right)
\gamma^{\mu    }
\left({{\bf \slash}\hskip-.65em\relax Q+
       {\bf \slash}\hskip-.5em\relax k_{1  }+m}\right)
\gamma^{\nu    }
\left({{\bf \slash}\hskip-.65em\relax Q+
       {\bf \slash}\hskip-.5em\relax k_{12 }+m}\right)
\gamma^{\lambda}
\left({{\bf \slash}\hskip-.65em\relax Q+
       {\bf \slash}\hskip-.5em\relax k_{123}+m}\right)
\gamma^{\sigma }\right]}
            {\left({ 2\,Q.k_{1  }   +k_{1  }^2}\right)
             \left({ 2\,Q.k_{12 }   +k_{12 }^2}\right)
             \left({ 2\,Q.k_{123}   +k_{123}^2}\right)}},
\end{equation}
%%%%%%%%%%%%%%%%%%%%%%%%%%%%%%%%%%%%%%%%%%%%%%%%%%%%%%%%%%%%
where $k_{12} =k_1 +k_2$, etc. In Eq. {\ref{eq1},
the $k^{0}_i$ are to be given (small)
imaginary parts. Different choices of signs for these imaginary parts
define the different branches of the analytically continued imaginary-time
amplitude \cite{Evans}. The explicit result for the trace
in Eq. (\ref{B1234}) is given in Eq. (\ref{trace}) of the Appendix.

We may alternatively write the sum in Eq. (\ref{eq1}) as
\begin{equation}\label{eq3}
\sum_{(ijkl)'} \left [
B_{(ijkl)}^{\mu \nu \lambda \sigma}(k_1, k_2, k_3;Q) ~+~
B_{(ijkl)}^{\mu \nu \lambda \sigma}(k_1, k_2, k_3;-Q) \right ],
\end{equation}
where the $(ijkl)'$ in the summation indicates that only even permutations of
$(1234)$ are to be included.
%%%%%%%%%%%%%%%%%%%%%%%%%% Fig .2 %%%%%%%%%%%%%%%%%%%%%%%%%%%%%%%%%%%%%%%%%%
\begin{figure}[t]
\setlength{\unitlength}{0.008000in}%
\begingroup\makeatletter\ifx\SetFigFont\undefined
% extract first six characters in \fmtname
\def\x#1#2#3#4#5#6#7\relax{\def\x{#1#2#3#4#5#6}}%
\expandafter\x\fmtname xxxxxx\relax \def\y{splain}%
\ifx\x\y   % LaTeX or SliTeX?
\gdef\SetFigFont#1#2#3{%
  \ifnum #1<17\tiny\else \ifnum #1<20\small\else
  \ifnum #1<24\normalsize\else \ifnum #1<29\large\else
  \ifnum #1<34\Large\else \ifnum #1<41\LARGE\else
     \huge\fi\fi\fi\fi\fi\fi
  \csname #3\endcsname}%
\else
\gdef\SetFigFont#1#2#3{\begingroup
  \count@#1\relax \ifnum 25<\count@\count@25\fi
  \def\x{\endgroup\@setsize\SetFigFont{#2pt}}%
  \expandafter\x
    \csname \romannumeral\the\count@ pt\expandafter\endcsname
    \csname @\romannumeral\the\count@ pt\endcsname
  \csname #3\endcsname}%
\fi
\fi\endgroup
\begin{picture}(720,204)(120,405)
%
%Feito por Fernando Tadeu C. Brandt em 30/6/94
%
\thicklines
\put(645,440){\vector( 1, 0){140}}
\put(785,440){\line( 1, 0){ 55}}
\put(120,440){\vector( 1, 0){ 65}}
\put(185,440){\vector( 1, 0){140}}
\put(325,440){\vector( 1, 0){160}}
\put(485,440){\vector( 1, 0){160}}
\multiput(240,560)(0.0,-4.1){30}{\makebox(0.9,1.3){\SetFigFont{10}{12}{rm}.}}
\multiput(400,560)(0.0,-4.1){30}{\makebox(0.9,1.3){\SetFigFont{10}{12}{rm}.}}
\multiput(560,560)(0.0,-4.1){30}{\makebox(0.9,1.3){\SetFigFont{10}{12}{rm}.}}
\multiput(720,560)(0.0,-4.1){30}{\makebox(0.9,1.3){\SetFigFont{10}{12}{rm}.}}
\put(170,410){\makebox(0,0)[lb]{\smash{\SetFigFont{12}{14.4}{it}Q}}}
\put(220,580){\makebox(0,0)[lb]{\smash{\SetFigFont{12}{14.4}{it}k}}}
\put(380,580){\makebox(0,0)[lb]{\smash{\SetFigFont{12}{14.4}{it}k}}}
\put(540,580){\makebox(0,0)[lb]{\smash{\SetFigFont{12}{14.4}{it}k}}}
\put(700,580){\makebox(0,0)[lb]{\smash{\SetFigFont{12}{14.4}{it}k}}}
\put(230,570){\makebox(0,0)[lb]{\smash{\SetFigFont{8}{8.0}{rm}1}}}
\put(390,570){\makebox(0,0)[lb]{\smash{\SetFigFont{8}{8.0}{rm}2}}}
\put(550,570){\makebox(0,0)[lb]{\smash{\SetFigFont{8}{8.0}{rm}3}}}
\put(710,570){\makebox(0,0)[lb]{\smash{\SetFigFont{8}{8.0}{rm}4}}}
\put(240,580){\makebox(0,0)[lb]{\smash{\SetFigFont{12}{14.4}{rm}, $\mu$}}}
\put(400,580){\makebox(0,0)[lb]{\smash{\SetFigFont{12}{14.4}{rm}, $\nu$}}}
\put(560,580){\makebox(0,0)[lb]{\smash{\SetFigFont{12}{14.4}{rm}, $\lambda$}}}
\put(720,580){\makebox(0,0)[lb]{\smash{\SetFigFont{12}{14.4}{rm}, $\sigma$}}}
\put(765,410){\makebox(0,0)[lb]{\smash{\SetFigFont{12}{14.4}{it}Q}}}
\put(295,410){\makebox(0,0)[lb]{\smash{\SetFigFont{12}{14.4}{it}Q}}}
\put(455,410){\makebox(0,0)[lb]{\smash{\SetFigFont{12}{14.4}{it}Q}}}
\put(615,410){\makebox(0,0)[lb]{\smash{\SetFigFont{12}{14.4}{it}Q}}}
\put(315,410){\makebox(0,0)[lb]{\smash{\SetFigFont{12}{14.4}{rm}+}}}
\put(475,410){\makebox(0,0)[lb]{\smash{\SetFigFont{12}{14.4}{rm}+}}}
\put(635,410){\makebox(0,0)[lb]{\smash{\SetFigFont{12}{14.4}{rm}+}}}
\put(335,410){\makebox(0,0)[lb]{\smash{\SetFigFont{12}{14.4}{it}k}}}
\put(495,410){\makebox(0,0)[lb]{\smash{\SetFigFont{12}{14.4}{it}k}}}
\put(655,410){\makebox(0,0)[lb]{\smash{\SetFigFont{12}{14.4}{it}k}}}
\put(345,400){\makebox(0,0)[lb]{\smash{\SetFigFont{8}{8.0}{rm}1}}}
\put(505,400){\makebox(0,0)[lb]{\smash{\SetFigFont{8}{8.0}{rm}12}}}
\put(665,400){\makebox(0,0)[lb]{\smash{\SetFigFont{8}{8.0}{rm}123}}}
\end{picture}

\nopagebreak
\bigskip
\caption[f2]{\label{f2}{One of the four diagrams corresponding to
           the first diagram in Fig. 1.
           Dotted lines represent photons, and
           solid lines stand for electrons.}}
\end{figure}
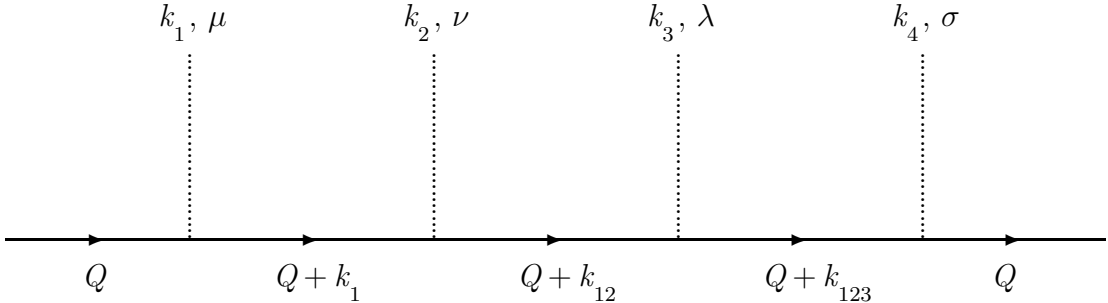
%%%%%%%%%%%%%%%%%%%%%%%%%%%%%%%%%%%%%%%%%%%%%%%%%%%%%%%%%%%%%%%%%%%%%%%%%%%

\section{The high-temperature behaviour}

\label{SEC3}
This section is included for completeness. Its conclusion is in fact
contained, as the abelian special case, in reference \cite{High-templimit}.
The straightforward way to get the high-temperature limit of
Eq. (\ref{eq1}) is to use the expansion
\begin{equation}\label{ExpDen}
(2Q.k +k^2)^{-1} =(2Q.k)^{-1}-k^2 (2Q.k)^{-2} + k^4 (2Q.k)^{-3}+\cdots
\end{equation}
in the denominators in
Eq. (\ref{B1234}), and to expand the numerator similarly in terms of
the $k_i$. One thus gets in
Eq. (\ref{B1234}) terms which are homogeneous in $Q$ of
degrees $1$, $0$, $-1$, $-2$. (The expansion
(\ref{ExpDen}) cannot be taken further
without introducing specious infra-red divergences.) The terms of odd
degree cancel in Eq. (\ref{eq3}).

The terms of degree $0$ would produce terms ${\cal O}(T^2)$ in (1).
However, such terms cancel in the sum in
Eq. (\ref{eq3}). As an example, take the terms
proportional to $Q^{\mu}Q^{\nu}Q^{\lambda}Q^{\sigma}$.
We find the following contributions (writing $K_i =Q.k_i$, etc):
\begin{equation}\label{cancelT2}
\begin{array}{ll}
&\displaystyle{\frac{k_1^2}{K_1^2}\left[
\frac{1}{K_{23}}\left(\frac{1}{K_2}+\frac{1}{K_3}\right)+
\frac{1}{K_{34}}\left(\frac{1}{K_3}+\frac{1}{K_4}\right)+
\frac{1}{K_{42}}\left(\frac{1}{K_4}+\frac{1}{K_2}\right)\right]+
(3~\hbox{cyclic perms.})}\\
&\displaystyle{
-\frac{k_{12}^2}{K_{12}^2}\left(\frac{1}{K_3}+\frac{1}{K_4}\right)
\left(\frac{1}{K_1}+\frac{1}{K_2}\right )+
(2~\hbox {cyclic perms. of 2,3,4})}\\
&\displaystyle{
=\frac{1}{K_1 K_2 K_3 K_4}
\left(k_1^2 +k_2^2 +k_3^2 +k_4^2-k_{12}^2-k_{13}^2-k_{14}^2\right)=0.}
\end{array}
\end{equation}

For the terms of degree $-2$, there is no such cancellation at the integrand
level in general. So, to examine possible $\log(T)$ terms, we have had to
resort
to an indirect argument. This we present in the next two sections.

\section{Absence of logarithmic temperature dependence}

\label{SEC4}

We begin the argument by considering the static case,
$k^{0i}=0,~i=1,\;2,\;3,\;4$.
This is simpler than the general case, because no analytic continuation
is necessary, and so we can write an explicit general formula for the
thermal box diagrams.
This has the form (unlike in Eq. (\ref{eq1}), we now refer to the complete
thermal amplitude, including the zero-temperature piece)
\begin{equation}\label{total}
T\sum_{Q^0=\pi iT(2n+1)}{1 \over (2\pi )^3}\int
{\rm d}^{3-\epsilon}{\bf Q}
F^{\mu \nu \lambda \sigma}({\bf k}_i, {\bf Q}, Q^0),
\end{equation}
where $n$ runs over the integers. For fixed $n$, the ${\bf Q}$-integral is IR
and UV finite (i.e. has no pole at $\epsilon =0$). For
\begin{equation}\label{HighT}
T\gg |{\bf k}_i|,~m,
\end{equation}
Eq. (\ref{total}) gives
\begin{equation}\label{HighTtotal}
\sum_{n} \left [ a(\epsilon)T^{-\epsilon}(2n+1)^{-1-\epsilon} + O(T^{-1-
\epsilon}
n^{-2-\epsilon}) \right ].
\end{equation}
If $a(0)$ is nonzero, this sum diverges; and the corresponding zero-temperature
Euclidean field theory (with $\int_{-i\infty}^{i\infty} dQ^0$
instead of the sum)would be UV divergent, which we know is not the case.
It follows that
\begin{equation}\label{UVdivergent}
a(\epsilon) \sim \epsilon b({\bf k}_i/m).
\end{equation}
We may  evaluate the sum in (\ref{total}), then let
$\epsilon \rightarrow 0$ and
finally let $T\rightarrow \infty$ to get a finite result.

As an example, let us consider the special case when $|{\bf k}^i| \ll m \ll T$.
Then we may write (\ref{total}) in the form
\begin{equation}\label{StaticSoft}
\int\frac{{\rm d}^3{\bf Q}}{(2\pi )^3}
\int\frac{{\rm d}z}{2\pi i} N(z)
\frac{N^{\mu \nu \lambda \sigma}({\bf k}_i=0,{\bf Q},z)}{(z^2 -q^2)^4},
\end{equation}
where $N^{\mu \nu \lambda \sigma}$ is the numerator of
$F^{\mu \nu \lambda \sigma}$.
First take the case $\mu =\nu =\lambda =\sigma =0$. Then all 3 graphs
in Fig. \ref{f1}, together with the ones obtained by charge conjugation,
contribute the same, and
\begin{equation}\label{F0000}
\begin{array}{ll}
N^{0000}&= {\rm tr}\left[(z+{\bf Q}\cdot{\bf \alpha})^4\right]\\
        &= 2\left[(z+q)^4+(z-q)^4\right],
\end{array}
\end{equation}
where ${\alpha^i}\equiv\gamma^{i}{\gamma^{0}}$. So (\ref{StaticSoft}) gives
\begin{equation}\label{StaticSoft0000}
\begin{array}{ll}\displaystyle{
\frac{6}{\pi^2}}\int_0^{\infty}{\rm d}q
&q^2\int
\displaystyle{\frac{{\rm d}z}{2\pi i}}N(z)
\left[\displaystyle{\frac{1}{(z-q)^4}}
     +\displaystyle{\frac{1}{(z+q)^4}}\right ]\\
&=\displaystyle{\frac{1}{\pi^2}}
\int_0^{\infty}{\rm d}q q^2 [N'''(q)+N'''(-q)]\\
&=-\displaystyle{\frac{4}{\pi^2}N(0)}
=-\displaystyle{\frac{2}{\pi^2}}.
\end{array}
\end{equation}

In this special case, a $\log(T)$ term would have shown itself as
\begin{equation}\label{IRintegral1}
\int^{\infty} \frac{{\rm d}q}{q} N(q).
\end{equation}
The IR divergence at the lower limit would have been a consequence of
using the expansion (\ref{ExpDen}).
But the presence of a term like (\ref{IRintegral1}) would have
correctly signalled the presence of a $\log(T)$ term. As it is we do not
find a term like (\ref{IRintegral1}) in (\ref{StaticSoft}).

The other components, $F^{00ij}$ etc, must vanish in the limit ${\bf k}_i
\rightarrow 0$ because of gauge invariance.

We now go from the static case to the general case, so far as $\log(T)$ terms
are concerned. Terms of degree $-2$ in $Q$ in the expansion of
(\ref{B1234}) using (\ref{ExpDen}) would produce an integral containing
\begin{equation}\label{IRintegral2}
\int_0^{\infty}{{\rm d}q \over Q^0}N(Q^0)
\end{equation}
(like (\ref{IRintegral1})), and thus a $\log(T)$ term.
%The angular integral in (\ref{eq1}) would then
%have an integrand which was
Setting $m=0$ in the angular integral in Eq. (\ref{eq1}), because mass terms
do not affect the $\log(T)$ contributions, we then obtain an integrand which
is
homogeneous of degree $-2$ in the vector
$\hat Q =(1,\hat {\bf Q})$. Such integrals have been studied in
\cite{High-templimit},
where it was shown that they could be expressed in the form
\begin{equation}\label{LorentzInv}
\sum_a R_a (k_i) M_a (k_i),
\end{equation}
where the $R_a$ are rational, Lorentz covariant tensors and the $M_a$
are a set of four
Lorentz invariant functions, with
\begin{equation}\label{LorentzInvFun}
M_1 = \Delta^{-\frac{1}{2}} \arctan \left [\frac{\Delta^\frac{1}{2}}{k_2.k_3}
\right ],
\end{equation}
where
\begin{equation}\label{Delta}
\Delta = k_2^2k_3^2 -(k_2.k_3)^2,
\end{equation}
and $M_2, M_3$ are defined similarly by cyclic permutation of $(1,2,3)$.
Finally $M_4 =1$. The important point about (\ref{LorentzInv})
is its Lorentz covariance,
in spite of the fact that the angular integral in (\ref{eq1})
is not generally a Lorentz invariant process (except when the integrand
is homogeneous in $\hat Q$ of degree $-2$).

Because of the importance for our proof of this result about Lorentz
invariance, we here insert a simple argument for this.
%Consider equation (\ref{eq1}) where we can set for our purpose $m=0$,
%since the masses do not contribute to the $\log(T)$ terms. Then we
%obtain an angular integral of the form
Consider the part of the angular integral in Eq. (\ref{eq1}) contributing
to $\log(T)$ terms, which has the form
\begin{equation}\label{AngInt1}
\int {\rm d} \Omega B(k_i, \hat Q).
\end{equation}
Because of homogeneity property, we can write this as
\begin{equation}\label{AngInt2}
\int d\Omega |{\bf Q}|^2[B(k_i, Q)]_{Q^0=|{\bf Q}|},
\end{equation}
for any value of $|{\bf Q}|$. In particular, we may fix ${k}.{Q}=1$
where $k$ is an arbitrary combination of the $k_i$. Thus we get
\begin{equation}\label{AngInt3}
2\int d^4Q \delta(Q^2)\theta (Q^0) \delta (k.Q-1) B(k_i, Q).
\end{equation}
This final form shows that the integral is a covariant function of the $k_i$.
But the original form shows that it also independent of the particular
choice of $k$. Thus the integral is covariant, as claimed.

We now return to the form (\ref{LorentzInv}).
Clearly the functions defined in (\ref{LorentzInvFun})
have branch cuts at $k_i^2 =0$, and the
coefficient rational functions $R_a$ in (\ref{LorentzInv})
have poles where subsets of the
$k_i$ are linearly dependent. In order to avoid these complications, we
now assume that all the $k_i^0$ are pure imaginary, i.e. we are dealing
with (4-dimensional) Euclidean field theory. If we can show that there is
no $\log(T)$ term in this case, then there can be no such term when we
analytically continue further towards the real $k_i^0$ axes.

Now suppose that there was a $\log(T)$ term in the non-static
case; so that not all
the $R_a$ in (\ref{LorentzInv}) were zero.
We shall show that this would contradict
our previous conclusion that there was no $\log(T)$ term in the static case.
Since (\ref{LorentzInv}) is covariant under rotations in 4-dimensional
Euclidean space, we may go to a coordinate system in which the (imaginary)
time axis is perpendicular to each of the $k_i$.
%Define
%\begin{equation}\label{nDirection}
%n_{\lambda} = \epsilon_{\lambda \mu \nu \sigma}k_1^{\mu}k_2^{\nu}k_3^{\sigma}.
%\end{equation}
This is just the static case ($k_i^0=0$). Thus the nonvanishing of
(\ref{LorentzInv})
would contradict what we know about the absence of
$\log(T)$ terms in that case.
This completes our argument about the general absence of $\log(T)$ terms.

\section{Discussion}

\label{SEC5}

In order to inquire further the absence of
$\log(T)$  terms in the box diagram,
we now briefly consider the photon self-energy in two-dimensional
QED, which is
similar in the sense that there is no UV divergence at zero-temperature.
Of course, the main difference
is the role of the axial anomaly in two-dimensional
QED \cite{2dim}, which has no counterpart in the four-dimensional box diagram.

Since this model is much simpler, we can give an explicit form for the
self-energy at infinite temperature. By using a method similar to that
in Sec. \ref{SEC4}, the high-temperature limit is found simply by setting
\begin{equation}\label{LimN}
N(z) \rightarrow \frac{1}{2}
\end{equation}
in the analogue of Eq. (\ref{StaticSoft}), evaluated for general
momenta $k_\mu$.
Then the result for the complete self-energy at infinite temperature is
\begin{equation}\label{Self2D}
\frac{e^2}{\pi}\left [\eta^{\mu \nu}-\frac{k^{\mu}k^{\nu}}{k^2}\right ].
\end{equation}
Note that the zero-temperature result is \cite{2dimT0}
\begin{equation}\label{Self2DT0}
\frac{e^2}{\pi}
\left [1+\frac{4m^2}{k^2 R}
       \arctan\left(R\right)\right]
\left[\eta^{\mu \nu}-\frac{k^{\mu}k^{\nu}}{k^2} \right],
\end{equation}
where $k^2<4m^2$ and $R=[4m^2 /k^2-1]^{1/2}$.
Thus the effect of the additional
thermal contributions (at high $T$) is simply to cancel the $m^2$ term,
leaving just the anomaly term behind.

An interesting question concerns the thermal behavior of QED Green functions
with $n>4$ external photon lines (for $n$-odd, these vanish identically
by charge conjugation). By power counting, these functions
are UV finite at zero temperature. Hence, using arguments like those in
Sec. \ref{SEC4}, we see that in the static case the $\log(T)$ terms
will be absent. Furthermore, from the work in references
\cite{High-templimit,BraatenPisarski}, it follows that the $T^2$
contributions will cancel out. Therefore, we conclude that in the
static case these Green functions will have a finite limit as
$T\rightarrow\infty$. Whether this continues to be true in general
still remains to be investigated.

\acknowledgements{F.T.B. and J.F. thank to ${\rm CNP_Q}$, Brasil, for a grant.}
\newpage

\appendix
\section{}

In this appendix  we present another explicit result for the high-temperature
behavior of the nonlinear electromagnetic interaction. Since in Sec. \ref{SEC4}
we have shown that there is no $\log(T)$ contribution and computed the
static case, it would be also interesting to know the
explicit expression for the {\it long wavelength limit}.

Our starting point is Eq. (\ref{B1234}). The computation
of the trace is a tedious but straightforward task which
can be accomplished using a computer program for symbolic manipulations.
The result is
\begin{displaymath}
{
\begin{array}{ll}
4\left(\right.
&
  k_{1}^{\sigma}\,k_{12}^{\lambda}\,k_{123}^{\nu}\,Q^{\mu} -
  k_{1}^{\lambda}\,k_{12}^{\sigma}\,k_{123}^{\nu}\,Q^{\mu} +
  k_{1}^{\sigma}\,k_{12}^{\nu}\,k_{123}^{\lambda}\,Q^{\mu} +
  k_{1}^{\nu}\,k_{12}^{\sigma}\,k_{123}^{\lambda}\,Q^{\mu} +\\
&
  k_{1}^{\lambda}\,k_{12}^{\nu}\,k_{123}^{\sigma}\,Q^{\mu} +
  k_{1}^{\nu}\,k_{12}^{\lambda}\,k_{123}^{\sigma}\,Q^{\mu} -
  k_{1}^{\sigma}\,k_{12}^{\lambda}\,k_{123}^{\mu}\,Q^{\nu} +
  k_{1}^{\lambda}\,k_{12}^{\sigma}\,k_{123}^{\mu}\,Q^{\nu} -\\
&
  k_{1}^{\sigma}\,k_{12}^{\mu}\,k_{123}^{\lambda}\,Q^{\nu} +
  k_{1}^{\mu}\,k_{12}^{\sigma}\,k_{123}^{\lambda}\,Q^{\nu} -
  k_{1}^{\lambda}\,k_{12}^{\mu}\,k_{123}^{\sigma}\,Q^{\nu} +
  k_{1}^{\mu}\,k_{12}^{\lambda}\,k_{123}^{\sigma}\,Q^{\nu} +\\
&
  2\,k_{12}^{\sigma}\,k_{123}^{\lambda}\,Q^{\mu}\,Q^{\nu} +
  2\,k_{12}^{\lambda}\,k_{123}^{\sigma}\,Q^{\mu}\,Q^{\nu} -
  k_{1}^{\sigma}\,k_{12}^{\nu}\,k_{123}^{\mu}\,Q^{\lambda} -
  k_{1}^{\nu}\,k_{12}^{\sigma}\,k_{123}^{\mu}\,Q^{\lambda} +\\
&
  k_{1}^{\sigma}\,k_{12}^{\mu}\,k_{123}^{\nu}\,Q^{\lambda} -
  k_{1}^{\mu}\,k_{12}^{\sigma}\,k_{123}^{\nu}\,Q^{\lambda} +
  k_{1}^{\nu}\,k_{12}^{\mu}\,k_{123}^{\sigma}\,Q^{\lambda} +
  k_{1}^{\mu}\,k_{12}^{\nu}\,k_{123}^{\sigma}\,Q^{\lambda} +\\
&
  2\,k_{1}^{\sigma}\,k_{123}^{\nu}\,Q^{\mu}\,Q^{\lambda} -
  2\,k_{12}^{\sigma}\,k_{123}^{\nu}\,Q^{\mu}\,Q^{\lambda} +
  2\,k_{1}^{\nu}\,k_{123}^{\sigma}\,Q^{\mu}\,Q^{\lambda} +
  2\,k_{12}^{\nu}\,k_{123}^{\sigma}\,Q^{\mu}\,Q^{\lambda} -\\
&
  2\,k_{1}^{\sigma}\,k_{123}^{\mu}\,Q^{\nu}\,Q^{\lambda} +
  2\,k_{1}^{\mu}\,k_{123}^{\sigma}\,Q^{\nu}\,Q^{\lambda} +
  4\,k_{123}^{\sigma}\,Q^{\mu}\,Q^{\nu}\,Q^{\lambda} +
  k_{1}^{\lambda}\,k_{12}^{\nu}\,k_{123}^{\mu}\,Q^{\sigma} +\\
&
  k_{1}^{\nu}\,k_{12}^{\lambda}\,k_{123}^{\mu}\,Q^{\sigma} -
  k_{1}^{\lambda}\,k_{12}^{\mu}\,k_{123}^{\nu}\,Q^{\sigma} +
  k_{1}^{\mu}\,k_{12}^{\lambda}\,k_{123}^{\nu}\,Q^{\sigma} +
  k_{1}^{\nu}\,k_{12}^{\mu}\,k_{123}^{\lambda}\,Q^{\sigma} +\\
&
  k_{1}^{\mu}\,k_{12}^{\nu}\,k_{123}^{\lambda}\,Q^{\sigma} +
  2\,k_{1}^{\lambda}\,k_{12}^{\nu}\,Q^{\mu}\,Q^{\sigma} +
  2\,k_{1}^{\nu}\,k_{12}^{\lambda}\,Q^{\mu}\,Q^{\sigma} -
  2\,k_{1}^{\lambda}\,k_{123}^{\nu}\,Q^{\mu}\,Q^{\sigma} +\\
&
  2\,k_{12}^{\lambda}\,k_{123}^{\nu}\,Q^{\mu}\,Q^{\sigma} +
  2\,k_{1}^{\nu}\,k_{123}^{\lambda}\,Q^{\mu}\,Q^{\sigma} +
  2\,k_{12}^{\nu}\,k_{123}^{\lambda}\,Q^{\mu}\,Q^{\sigma} -
  2\,k_{1}^{\lambda}\,k_{12}^{\mu}\,Q^{\nu}\,Q^{\sigma} +\\
&
  2\,k_{1}^{\mu}\,k_{12}^{\lambda}\,Q^{\nu}\,Q^{\sigma} +
  2\,k_{1}^{\lambda}\,k_{123}^{\mu}\,Q^{\nu}\,Q^{\sigma} +
  2\,k_{1}^{\mu}\,k_{123}^{\lambda}\,Q^{\nu}\,Q^{\sigma} +
  4\,k_{12}^{\lambda}\,Q^{\mu}\,Q^{\nu}\,Q^{\sigma} +\\
&
  4\,k_{123}^{\lambda}\,Q^{\mu}\,Q^{\nu}\,Q^{\sigma} +
  2\,k_{1}^{\nu}\,k_{12}^{\mu}\,Q^{\lambda}\,Q^{\sigma} +
  2\,k_{1}^{\mu}\,k_{12}^{\nu}\,Q^{\lambda}\,Q^{\sigma} +
  4\,k_{1}^{\nu}\,Q^{\mu}\,Q^{\lambda}\,Q^{\sigma} +\\
&
  4\,k_{12}^{\nu}\,Q^{\mu}\,Q^{\lambda}\,Q^{\sigma} +
  4\,k_{1}^{\mu}\,Q^{\nu}\,Q^{\lambda}\,Q^{\sigma} +
  8\,Q^{\mu}\,Q^{\nu}\,Q^{\lambda}\,Q^{\sigma} +
  k_{1}^{\sigma}\,k_{12}^{\lambda}\,k_{123} \cdot Q\,\eta^{\mu  \nu} -\\
&
  k_{1}^{\lambda}\,k_{12}^{\sigma}\,k_{123} \cdot Q\,\eta^{\mu  \nu} +
  k_{1}^{\sigma}\,k_{12} \cdot Q\,k_{123}^{\lambda}\,\eta^{\mu  \nu} -
  k_{1} \cdot Q\,k_{12}^{\sigma}\,k_{123}^{\lambda}\,\eta^{\mu  \nu} +
  k_{1}^{\lambda}\,k_{12} \cdot Q\,k_{123}^{\sigma}\,\eta^{\mu  \nu} -\\
&
  k_{1} \cdot Q\,k_{12}^{\lambda}\,k_{123}^{\sigma}\,\eta^{\mu  \nu} -
  k_{1}^{\sigma}\,k_{12} \cdot k_{123}\,Q^{\lambda}\,\eta^{\mu  \nu} +
  k_{1} \cdot k_{123}\,k_{12}^{\sigma}\,Q^{\lambda}\,\eta^{\mu  \nu} -
  k_{1} \cdot k_{12}\,k_{123}^{\sigma}\,Q^{\lambda}\,\eta^{\mu  \nu} -\\
&
  2\,k_{1} \cdot Q\,k_{123}^{\sigma}\,Q^{\lambda}\,\eta^{\mu  \nu} +
  k_{1}^{\lambda}\,k_{12} \cdot k_{123}\,Q^{\sigma}\,\eta^{\mu  \nu} +
  2\,k_{1}^{\lambda}\,k_{12} \cdot Q\,Q^{\sigma}\,\eta^{\mu  \nu} -
  k_{1} \cdot k_{123}\,k_{12}^{\lambda}\,Q^{\sigma}\,\eta^{\mu  \nu} -\\
&
  2\,k_{1} \cdot Q\,k_{12}^{\lambda}\,Q^{\sigma}\,\eta^{\mu  \nu} -
  k_{1} \cdot k_{12}\,k_{123}^{\lambda}\,Q^{\sigma}\,\eta^{\mu  \nu} -
  2\,k_{1} \cdot Q\,k_{123}^{\lambda}\,Q^{\sigma}\,\eta^{\mu  \nu} -
  2\,k_{1} \cdot k_{12}\,Q^{\lambda}\,Q^{\sigma}\,\eta^{\mu  \nu} -\\
&
  4\,k_{1} \cdot Q\,Q^{\lambda}\,Q^{\sigma}\,\eta^{\mu  \nu} +
  k_{1}^{\sigma}\,k_{12}^{\nu}\,k_{123} \cdot Q\,\eta^{\mu  \lambda} +
  k_{1}^{\nu}\,k_{12}^{\sigma}\,k_{123} \cdot Q\,\eta^{\mu  \lambda} -
  k_{1}^{\sigma}\,k_{12} \cdot Q\,k_{123}^{\nu}\,\eta^{\mu  \lambda} +\\
&
  k_{1} \cdot Q\,k_{12}^{\sigma}\,k_{123}^{\nu}\,\eta^{\mu  \lambda} -
  k_{1}^{\nu}\,k_{12} \cdot Q\,k_{123}^{\sigma}\,\eta^{\mu  \lambda} -
  k_{1} \cdot Q\,k_{12}^{\nu}\,k_{123}^{\sigma}\,\eta^{\mu  \lambda} +
  k_{1}^{\sigma}\,k_{12} \cdot k_{123}\,Q^{\nu}\,\eta^{\mu  \lambda} -\\
\end{array}}
\end{displaymath}
\begin{equation}\label{trace}
{
\begin{array}{ll}
&
  k_{1} \cdot k_{123}\,k_{12}^{\sigma}\,Q^{\nu}\,\eta^{\mu  \lambda} +
  2\,k_{1}^{\sigma}\,k_{123} \cdot Q\,Q^{\nu}\,\eta^{\mu  \lambda} +
  k_{1} \cdot k_{12}\,k_{123}^{\sigma}\,Q^{\nu}\,\eta^{\mu  \lambda} -
  k_{1}^{\nu}\,k_{12} \cdot k_{123}\,Q^{\sigma}\,\eta^{\mu  \lambda} -\\
&
  2\,k_{1}^{\nu}\,k_{12} \cdot Q\,Q^{\sigma}\,\eta^{\mu  \lambda} -
  k_{1} \cdot k_{123}\,k_{12}^{\nu}\,Q^{\sigma}\,\eta^{\mu  \lambda} -
  2\,k_{1} \cdot Q\,k_{12}^{\nu}\,Q^{\sigma}\,\eta^{\mu  \lambda} +
  k_{1} \cdot k_{12}\,k_{123}^{\nu}\,Q^{\sigma}\,\eta^{\mu  \lambda} +\\
&
  2\,k_{1} \cdot Q\,k_{123}^{\nu}\,Q^{\sigma}\,\eta^{\mu  \lambda} +
  2\,k_{1} \cdot k_{12}\,Q^{\nu}\,Q^{\sigma}\,\eta^{\mu  \lambda} -
  2\,k_{1} \cdot k_{123}\,Q^{\nu}\,Q^{\sigma}\,\eta^{\mu  \lambda} -
  k_{1}^{\lambda}\,k_{12}^{\nu}\,k_{123} \cdot Q\,\eta^{\mu  \sigma} -\\
&
  k_{1}^{\nu}\,k_{12}^{\lambda}\,k_{123} \cdot Q\,\eta^{\mu  \sigma} +
  k_{1}^{\lambda}\,k_{12} \cdot Q\,k_{123}^{\nu}\,\eta^{\mu  \sigma} -
  k_{1} \cdot Q\,k_{12}^{\lambda}\,k_{123}^{\nu}\,\eta^{\mu  \sigma} -
  k_{1}^{\nu}\,k_{12} \cdot Q\,k_{123}^{\lambda}\,\eta^{\mu  \sigma} -\\
&
  k_{1} \cdot Q\,k_{12}^{\nu}\,k_{123}^{\lambda}\,\eta^{\mu  \sigma} -
  k_{1}^{\lambda}\,k_{12} \cdot k_{123}\,Q^{\nu}\,\eta^{\mu  \sigma} +
  k_{1} \cdot k_{123}\,k_{12}^{\lambda}\,Q^{\nu}\,\eta^{\mu  \sigma} -
  2\,k_{1}^{\lambda}\,k_{123} \cdot Q\,Q^{\nu}\,\eta^{\mu  \sigma} +\\
&
  k_{1} \cdot k_{12}\,k_{123}^{\lambda}\,Q^{\nu}\,\eta^{\mu  \sigma} +
  k_{1}^{\nu}\,k_{12} \cdot k_{123}\,Q^{\lambda}\,\eta^{\mu  \sigma} +
  k_{1} \cdot k_{123}\,k_{12}^{\nu}\,Q^{\lambda}\,\eta^{\mu  \sigma} -
  k_{1} \cdot k_{12}\,k_{123}^{\nu}\,Q^{\lambda}\,\eta^{\mu  \sigma} -\\
&
  2\,k_{1} \cdot Q\,k_{123}^{\nu}\,Q^{\lambda}\,\eta^{\mu  \sigma} +
  2\,k_{1} \cdot k_{123}\,Q^{\nu}\,Q^{\lambda}\,\eta^{\mu  \sigma} -
  k_{1}^{\sigma}\,k_{12}^{\mu}\,k_{123} \cdot Q\,\eta^{\nu  \lambda} +
  k_{1}^{\mu}\,k_{12}^{\sigma}\,k_{123} \cdot Q\,\eta^{\nu  \lambda} +\\
&
  k_{1}^{\sigma}\,k_{12} \cdot Q\,k_{123}^{\mu}\,\eta^{\nu  \lambda} -
  k_{1} \cdot Q\,k_{12}^{\sigma}\,k_{123}^{\mu}\,\eta^{\nu  \lambda} -
  k_{1}^{\mu}\,k_{12} \cdot Q\,k_{123}^{\sigma}\,\eta^{\nu  \lambda} +
  k_{1} \cdot Q\,k_{12}^{\mu}\,k_{123}^{\sigma}\,\eta^{\nu  \lambda} -\\
&
  k_{1}^{\sigma}\,k_{12} \cdot k_{123}\,Q^{\mu}\,\eta^{\nu  \lambda} +
  k_{1} \cdot k_{123}\,k_{12}^{\sigma}\,Q^{\mu}\,\eta^{\nu  \lambda} -
  2\,k_{1}^{\sigma}\,k_{123} \cdot Q\,Q^{\mu}\,\eta^{\nu  \lambda} +
  2\,k_{12}^{\sigma}\,k_{123} \cdot Q\,Q^{\mu}\,\eta^{\nu  \lambda} -\\
&
  k_{1} \cdot k_{12}\,k_{123}^{\sigma}\,Q^{\mu}\,\eta^{\nu  \lambda} -
  2\,k_{12} \cdot Q\,k_{123}^{\sigma}\,Q^{\mu}\,\eta^{\nu  \lambda} -
  k_{1}^{\mu}\,k_{12} \cdot k_{123}\,Q^{\sigma}\,\eta^{\nu  \lambda} -
  2\,k_{1}^{\mu}\,k_{12} \cdot Q\,Q^{\sigma}\,\eta^{\nu  \lambda} +\\
&
  k_{1} \cdot k_{123}\,k_{12}^{\mu}\,Q^{\sigma}\,\eta^{\nu  \lambda} +
  2\,k_{1} \cdot Q\,k_{12}^{\mu}\,Q^{\sigma}\,\eta^{\nu  \lambda} -
  k_{1} \cdot k_{12}\,k_{123}^{\mu}\,Q^{\sigma}\,\eta^{\nu  \lambda} -
  2\,k_{1} \cdot Q\,k_{123}^{\mu}\,Q^{\sigma}\,\eta^{\nu  \lambda} -\\
&
  2\,k_{1} \cdot k_{12}\,Q^{\mu}\,Q^{\sigma}\,\eta^{\nu  \lambda} +
  2\,k_{1} \cdot k_{123}\,Q^{\mu}\,Q^{\sigma}\,\eta^{\nu  \lambda} -
  2\,k_{12} \cdot k_{123}\,Q^{\mu}\,Q^{\sigma}\,\eta^{\nu  \lambda} -
  4\,k_{12} \cdot Q\,Q^{\mu}\,Q^{\sigma}\,\eta^{\nu  \lambda} +\\
&
  k_{1} \cdot Q\,k_{12} \cdot k_{123}\,\eta^{\mu  \sigma}\,
   \eta^{\nu  \lambda} - k_{1} \cdot k_{123}\,k_{12} \cdot Q\,
   \eta^{\mu  \sigma}\,\eta^{\nu  \lambda} +
  k_{1} \cdot k_{12}\,k_{123} \cdot Q\,\eta^{\mu  \sigma}\,
   \eta^{\nu  \lambda} +  2\,k_{1} \cdot Q\,k_{123} \cdot Q\,
   \eta^{\mu  \sigma}\,\eta^{\nu  \lambda} +\\
&
  k_{1}^{\lambda}\,k_{12}^{\mu}\,k_{123} \cdot Q\,\eta^{\nu  \sigma} -
  k_{1}^{\mu}\,k_{12}^{\lambda}\,k_{123} \cdot Q\,\eta^{\nu  \sigma} -
  k_{1}^{\lambda}\,k_{12} \cdot Q\,k_{123}^{\mu}\,\eta^{\nu  \sigma} +
  k_{1} \cdot Q\,k_{12}^{\lambda}\,k_{123}^{\mu}\,\eta^{\nu  \sigma} -\\
&
  k_{1}^{\mu}\,k_{12} \cdot Q\,k_{123}^{\lambda}\,\eta^{\nu  \sigma} +
  k_{1} \cdot Q\,k_{12}^{\mu}\,k_{123}^{\lambda}\,\eta^{\nu  \sigma} +
  k_{1}^{\lambda}\,k_{12} \cdot k_{123}\,Q^{\mu}\,\eta^{\nu  \sigma} -
  k_{1} \cdot k_{123}\,k_{12}^{\lambda}\,Q^{\mu}\,\eta^{\nu  \sigma} +\\
&
  2\,k_{1}^{\lambda}\,k_{123} \cdot Q\,Q^{\mu}\,\eta^{\nu  \sigma} -
  2\,k_{12}^{\lambda}\,k_{123} \cdot Q\,Q^{\mu}\,\eta^{\nu  \sigma} -
  k_{1} \cdot k_{12}\,k_{123}^{\lambda}\,Q^{\mu}\,\eta^{\nu  \sigma} -
  2\,k_{12} \cdot Q\,k_{123}^{\lambda}\,Q^{\mu}\,\eta^{\nu  \sigma} +\\
&
  k_{1}^{\mu}\,k_{12} \cdot k_{123}\,Q^{\lambda}\,\eta^{\nu  \sigma} -
  k_{1} \cdot k_{123}\,k_{12}^{\mu}\,Q^{\lambda}\,\eta^{\nu  \sigma} +
  k_{1} \cdot k_{12}\,k_{123}^{\mu}\,Q^{\lambda}\,\eta^{\nu  \sigma} +
  2\,k_{1} \cdot Q\,k_{123}^{\mu}\,Q^{\lambda}\,\eta^{\nu  \sigma} -\\
&
  2\,k_{1} \cdot k_{123}\,Q^{\mu}\,Q^{\lambda}\,\eta^{\nu  \sigma} +
  2\,k_{12} \cdot k_{123}\,Q^{\mu}\,Q^{\lambda}\,\eta^{\nu  \sigma} -
  k_{1} \cdot Q\,k_{12} \cdot k_{123}\,\eta^{\mu  \lambda}\,
   \eta^{\nu  \sigma} +  k_{1} \cdot k_{123}\,k_{12} \cdot Q\,
   \eta^{\mu  \lambda}\,\eta^{\nu  \sigma} -\\
&
  k_{1} \cdot k_{12}\,k_{123} \cdot Q\,\eta^{\mu  \lambda}\,
   \eta^{\nu  \sigma} - 2\,k_{1} \cdot Q\,k_{123} \cdot Q\,
   \eta^{\mu  \lambda}\,\eta^{\nu  \sigma} -
  k_{1}^{\nu}\,k_{12}^{\mu}\,k_{123} \cdot Q\,\eta^{\lambda  \sigma} -
  k_{1}^{\mu}\,k_{12}^{\nu}\,k_{123} \cdot Q\,\eta^{\lambda  \sigma} +\\
&
  k_{1}^{\nu}\,k_{12} \cdot Q\,k_{123}^{\mu}\,\eta^{\lambda  \sigma} +
  k_{1} \cdot Q\,k_{12}^{\nu}\,k_{123}^{\mu}\,\eta^{\lambda  \sigma} +
  k_{1}^{\mu}\,k_{12} \cdot Q\,k_{123}^{\nu}\,\eta^{\lambda  \sigma} -
  k_{1} \cdot Q\,k_{12}^{\mu}\,k_{123}^{\nu}\,\eta^{\lambda  \sigma} -\\
&
  k_{1}^{\nu}\,k_{12} \cdot k_{123}\,Q^{\mu}\,\eta^{\lambda  \sigma} -
  k_{1} \cdot k_{123}\,k_{12}^{\nu}\,Q^{\mu}\,\eta^{\lambda  \sigma} -
  2\,k_{1}^{\nu}\,k_{123} \cdot Q\,Q^{\mu}\,\eta^{\lambda  \sigma} -
  2\,k_{12}^{\nu}\,k_{123} \cdot Q\,Q^{\mu}\,\eta^{\lambda  \sigma} +\\
&
  k_{1} \cdot k_{12}\,k_{123}^{\nu}\,Q^{\mu}\,\eta^{\lambda  \sigma} +
  2\,k_{12} \cdot Q\,k_{123}^{\nu}\,Q^{\mu}\,\eta^{\lambda  \sigma} -
  k_{1}^{\mu}\,k_{12} \cdot k_{123}\,Q^{\nu}\,\eta^{\lambda  \sigma} +
  k_{1} \cdot k_{123}\,k_{12}^{\mu}\,Q^{\nu}\,\eta^{\lambda  \sigma} -\\
&
  2\,k_{1}^{\mu}\,k_{123} \cdot Q\,Q^{\nu}\,\eta^{\lambda  \sigma} -
  k_{1} \cdot k_{12}\,k_{123}^{\mu}\,Q^{\nu}\,\eta^{\lambda  \sigma} -
  2\,k_{12} \cdot k_{123}\,Q^{\mu}\,Q^{\nu}\,\eta^{\lambda  \sigma} -
  4\,k_{123} \cdot Q\,Q^{\mu}\,Q^{\nu}\,\eta^{\lambda  \sigma}+\\
&
  k_{1} \cdot Q\,k_{12} \cdot k_{123}\,\eta^{\mu  \nu}\,
   \eta^{\lambda  \sigma} - k_{1} \cdot k_{123}\,k_{12} \cdot Q\,
   \eta^{\mu  \nu}\,\eta^{\lambda  \sigma} +
  k_{1} \cdot k_{12}\,k_{123} \cdot Q\,\eta^{\mu  \nu}\,
   \eta^{\lambda  \sigma} +  2\,k_{1} \cdot Q\,k_{123} \cdot Q \,
   \eta^{\mu  \nu}\,\eta^{\lambda  \sigma}
\left.\right)
\\
\end{array}}
\end{equation}

In the long wavelength limit all wave vectors are set equal to zero and
Eq. (\ref{trace}) simplifies considerably. The denominators in Eq.
(\ref{B1234}) will not depend on ${\bf Q}$ and the
angular integrals in (\ref{eq1}) can be easily computed. Using
partial fractions decomposition the result can be expressed in terms of
integrals like
\begin{equation}\label{parfrac}
\int_0^\infty \frac{{\rm d} q}{\exp(q/T)+1}\frac{1}{q\pm s},
\end{equation}
where $s$ a linear function of the external frequencies. In the limit
$T\rightarrow\infty$ we can replace each individual integral by
\begin{equation}\label{long_wave_int}
\int_0^T \frac{{\rm d} q}{2}\frac{1}{q\pm s}=
\frac{1}{2}\left(\log(T)-\log(\pm s)+{\cal O}(1/T)\right).
\end{equation}
When we add together all the contributions, we obtain the result
\begin{equation}
\frac{e^4}{15\pi^2}
\left(A(\omega_1,\omega_2,\omega_3,\omega_4) \delta^{ij} \delta^{kl}+
      A(\omega_1,\omega_3,\omega_2,\omega_4) \delta^{ik} \delta^{jl}+
      A(\omega_1,\omega_4,\omega_2,\omega_3) \delta^{il} \delta^{jk}\right),
\end{equation}
where $\omega_1+\omega_2+\omega_3+\omega_4=0$ and the indices
$(i,\;j,\;k,\;l)$ are in correspondence with $(\mu,\;\nu,\;\lambda,\;\sigma)$.
The function $A(\omega_1,\omega_2,\omega_3,\omega_4)$ is a combination
of rational functions of the frequencies times expressions like
(\ref{long_wave_int}), and it is symmetric under the permutations
$\omega_1\leftrightarrow\omega_2$, $\omega_3\leftrightarrow\omega_4$
and $(\omega_1,\, \omega_2)\leftrightarrow(\omega_3,\, \omega_4)$.

 From (\ref{long_wave_int}) we can see that
the function $A(\omega_1,\omega_2,\omega_3,\omega_4)$ grows not faster
than $\log(T)$ in the high temperature limit.
This is just a special case of the result in
Sec. \ref{SEC3}, where we have shown that the cancellation of contributions
proportional to $T^2$ occurs at the integrand level. We also expect from
Sec. \ref{SEC4} that the $\log(T)$ contributions should vanish. Indeed,
we have obtained that the coefficient of $\log(T)$ is proportional to
the following vanishing combination of rational functions
\begin{equation}\label{zero}
\begin{array}{lll}
&
c(\omega_1,\omega_2,\omega_3,\omega_4)+
c(\omega_2,\omega_1,\omega_3,\omega_4)+
c(\omega_3,\omega_4,\omega_1,\omega_2)+
c(\omega_4,\omega_3,\omega_1,\omega_2)+
\\&
d(\omega_1,\omega_2,\omega_3,\omega_4)+
d(\omega_2,\omega_1,\omega_3,\omega_4)+
e(\omega_1,\omega_2,\omega_3,\omega_4)=0
\end{array}
\end{equation}
where
\begin{mathletters}\label{Cs}
\begin{equation}
%\begin{array}{ll}
\label{ca}
%&
c(\omega_1,\omega_2,\omega_3,\omega_4)=
%\\&
%\displaystyle
{{{{{\omega_1}}^2}\,\left[ 10\,{{{\omega_2}}^2}\,{\omega_3}\,{\omega_4} +
       \left( {\omega_1} + 5\,{\omega_2} \right) \,
        \left( {\omega_1}\,{\omega_2}\,{\omega_{34}} +
          {\omega_1}\,{\omega_3}\,{\omega_4} \right)  \right] }\over
   {{\omega_{12}}\,{\omega_{13}}\,{\omega_2}\,{\omega_{23}}
   \,{\omega_3}\,{\omega_4}}},
%\end{array}
\end{equation}
\begin{equation}
%\begin{array}{ll}
d(\omega_1,\omega_2,\omega_3,\omega_4)=
%\\&
%\displaystyle
{{{{{\omega_{13}}}^2}\,
\left( {{{\omega_1}}^2} + 5\,{\omega_1}\,{\omega_2} -
3\,{\omega_1}\,{\omega_3} -
5\,{\omega_2}\,{\omega_3} - 4\,{{{\omega_3}}^2} \right) }\over
{{\omega_1}\,{\omega_2}\,{\omega_3}\,{\omega_4}}}
%\end{array}
\end{equation}
and
\begin{equation}
e(\omega_1,\omega_2,\omega_3,\omega_4)=
\frac{\left(\omega_{12}\;\omega_{34}\right)^2}
{\omega_1\;\omega_2\;\omega_3\;\omega_4}.
\end{equation}
\end{mathletters}
Finally, the function $A(\omega_1,\omega_2,\omega_3,\omega_4)$ is given by
\begin{equation}\label{As}
\begin{array}{lll}
A(\omega_1,\omega_2,\omega_3,\omega_4)=&

c(\omega_1,\omega_2,\omega_3,\omega_4)
\log({-\frac{\omega_1^2}{\omega_{12}\omega_{34}}})+

c(\omega_2,\omega_1,\omega_3,\omega_4)
\log({-\frac{\omega_2^2}{\omega_{12}\omega_{34}}})+\\&

c(\omega_3,\omega_4,\omega_1,\omega_2)
\log({-\frac{\omega_3^2}{\omega_{12}\omega_{34}}})+

c(\omega_4,\omega_3,\omega_1,\omega_2)
\log({-\frac{\omega_4^2}{\omega_{12}\omega_{34}}})+\\&

d(\omega_1,\omega_2,\omega_3,\omega_4)
\log({\frac{\omega_{13}\omega_{24}}{\omega_{12}\omega_{34}}})+

d(\omega_2,\omega_1,\omega_3,\omega_4)
\log({\frac{\omega_{23}\omega_{14}}{\omega_{12}\omega_{34}}}).
\end{array}
\end{equation}
The symmetries of $A(\omega_1,\omega_2,\omega_3,\omega_4)$ under permutations
$\omega_1\leftrightarrow\omega_2$, $\omega_3\leftrightarrow\omega_4$
and $(\omega_1,\, \omega_2)\leftrightarrow(\omega_3,\, \omega_4)$,
can be easily verified using momentum conservation and Eqs. (\ref{Cs}).
One can also notice that $A(\omega_1,\omega_2,\omega_3,\omega_4)$ is real
when all the frequencies are real.


\begin{thebibliography}{99}

\bibitem{KN}
R. Karplus and M. Neuman, Phys Rev. {\bf 80}, 380 (1950);
{\bf 83}, 776 (1951).

\bibitem{DittrichLoewe}
W. Dittrich, Phys. Rev. D {\bf 19}, 2385 (1979);
M. Loewe and J. C. Rojas, Phys. Rev. D {\bf 46}, 2689 (1992).

\bibitem{Evans}
T. S. Evans, Nucl. Phys. {\bf B374}, 340 (1992).

\bibitem{Taylor}
J. C. Taylor, Phys. Rev. D {\bf 47}, 725 (1993).

\bibitem{TARRACH}
R. Tarrach, Phys Lett. {\bf 133B}, 259 (1983).

\bibitem{Barton}
G. Barton, Ann. Phys. {\bf 200}, 271 (1990).

\bibitem{FT}
J. Frenkel and J. C. Taylor, Nucl. Phys. {\bf B374}, 156 (1992).

\bibitem{High-templimit}
J. Frenkel and J. C. Taylor, Nucl. Phys. {\bf B334}, 199 (1990).

\bibitem{2dim}
R. Baier and E. Pilon, Z. Phys. C52, 339 (1991).

\bibitem{2dimT0}
C. Adam, R.A. Bertlmann and P. Hofer, Riv. Nuovo Cim. 16-no.8, 1 (1993).

\bibitem{BraatenPisarski}
E. Braaten and R. D. Pisarski, Nucl. Phys. {\bf B339}, 310 (1990).

\end{thebibliography}
\end{document}